\newcommand{\vdate}{November 1994}
\newcommand{\cernnr}{7516/94}
\newcommand{\beq}{\begin{equation}}
\newcommand{\eeq}{\end{equation}}
\newcommand{\beqn}{\begin{eqnarray}}
\newcommand{\eeqn}{\end{eqnarray}}

\newcommand{\opone}{{1\!\!\!\!\,\,1}}

\newcommand{\esd}{\cally}

\newcommand{\cC}{{\cal C}}
\newcommand{\cD}{{\cal D}}
\newcommand{\cE}{{\cal E}}
\newcommand{\cF}{{\cal F}}
\newcommand{\cH}{{\cal H}}

\renewcommand{\cC}{{\cally C}}
\renewcommand{\cD}{{\cally D}}
\renewcommand{\cE}{{\cally E}}
\renewcommand{\cF}{{\cally F}}
\renewcommand{\cH}{{\cally H}}

\renewcommand{\cC}{{\esd C}}
\renewcommand{\cD}{{\esd D}}
\renewcommand{\cE}{{\esd E}}
\renewcommand{\cF}{{\esd F}}
\renewcommand{\cH}{{\esd H}}

\newcommand{\gothn}{\gothi}

\newcommand{\gH}{{\gothn{H}}}
\newcommand{\gL}{{\gothn{L}}}
\newcommand{\gM}{{\gothn{M}}}
\newcommand{\sC}{{\Bbb C}}
\newcommand{\sR}{{\Bbb R}}
\newcommand{\sP}{{\Bbb P}}

\newcommand{\sZ}{{\Bbb Z}}
\newcommand{\sN}{{\Bbb N}}

\newcommand{\itemsym}{{$\scriptstyle\Box$}}

\newcommand{\defeq}{{\stackrel{def}{\displaystyle =}}}
\newcommand{\di}{{\mbox{d}}}

\documentstyle[12pt,epsfig]{article}

\catcode`\@=11

 \font\tenmsx=msam10 scaled \magstep1
 \font\sevenmsx=msam8
 \font\fivemsx=msam6
 \font\tenmsy=msbm10 scaled \magstep1
 \font\sevenmsy=msbm8
 \font\fivemsy=msbm6

\newfam\msxfam
\newfam\msyfam
\textfont\msxfam=\tenmsx  \scriptfont\msxfam=\sevenmsx
  \scriptscriptfont\msxfam=\fivemsx
\textfont\msyfam=\tenmsy  \scriptfont\msyfam=\sevenmsy
  \scriptscriptfont\msyfam=\fivemsy

\def\hexnumber@#1{\ifnum#1<10 \number#1\else
 \ifnum#1=10 A\else\ifnum#1=11 B\else\ifnum#1=12 C\else
 \ifnum#1=13 D\else\ifnum#1=14 E\else\ifnum#1=15 F\fi\fi\fi\fi\fi\fi\fi}

\def\msx@{\hexnumber@\msxfam}
\def\msy@{\hexnumber@\msyfam}
\mathchardef\boxdot="2\msx@00
\mathchardef\boxplus="2\msx@01
\mathchardef\boxtimes="2\msx@02
\mathchardef\square="0\msx@03
\mathchardef\blacksquare="0\msx@04
\mathchardef\centerdot="2\msx@05
\mathchardef\lozenge="0\msx@06
\mathchardef\blacklozenge="0\msx@07
\mathchardef\circlearrowright="3\msx@08
\mathchardef\circlearrowleft="3\msx@09
\mathchardef\rightleftharpoons="3\msx@0A
\mathchardef\leftrightharpoons="3\msx@0B
\mathchardef\boxminus="2\msx@0C
\mathchardef\Vdash="3\msx@0D
\mathchardef\Vvdash="3\msx@0E
\mathchardef\vDash="3\msx@0F
\mathchardef\twoheadrightarrow="3\msx@10
\mathchardef\twoheadleftarrow="3\msx@11
\mathchardef\leftleftarrows="3\msx@12
\mathchardef\rightrightarrows="3\msx@13
\mathchardef\upuparrows="3\msx@14
\mathchardef\downdownarrows="3\msx@15
\mathchardef\upharpoonright="3\msx@16

\mathchardef\downharpoonright="3\msx@17
\mathchardef\upharpoonleft="3\msx@18
\mathchardef\downharpoonleft="3\msx@19
\mathchardef\rightarrowtail="3\msx@1A
\mathchardef\leftarrowtail="3\msx@1B
\mathchardef\leftrightarrows="3\msx@1C
\mathchardef\rightleftarrows="3\msx@1D
\mathchardef\Lsh="3\msx@1E
\mathchardef\Rsh="3\msx@1F
\mathchardef\rightsquigarrow="3\msx@20
\mathchardef\leftrightsquigarrow="3\msx@21
\mathchardef\looparrowleft="3\msx@22
\mathchardef\looparrowright="3\msx@23
\mathchardef\circeq="3\msx@24
\mathchardef\succsim="3\msx@25
\mathchardef\gtrsim="3\msx@26
\mathchardef\gtrapprox="3\msx@27
\mathchardef\multimap="3\msx@28
\mathchardef\therefore="3\msx@29
\mathchardef\because="3\msx@2A
\mathchardef\doteqdot="3\msx@2B

\mathchardef\triangleq="3\msx@2C
\mathchardef\precsim="3\msx@2D
\mathchardef\lesssim="3\msx@2E
\mathchardef\lessapprox="3\msx@2F
\mathchardef\eqslantless="3\msx@30
\mathchardef\eqslantgtr="3\msx@31
\mathchardef\curlyeqprec="3\msx@32
\mathchardef\curlyeqsucc="3\msx@33
\mathchardef\preccurlyeq="3\msx@34
\mathchardef\leqq="3\msx@35
\mathchardef\leqslant="3\msx@36
\mathchardef\lessgtr="3\msx@37
\mathchardef\backprime="0\msx@38
\mathchardef\risingdotseq="3\msx@3A
\mathchardef\fallingdotseq="3\msx@3B
\mathchardef\succcurlyeq="3\msx@3C
\mathchardef\geqq="3\msx@3D
\mathchardef\geqslant="3\msx@3E
\mathchardef\gtrless="3\msx@3F
\mathchardef\sqsubset="3\msx@40
\mathchardef\sqsupset="3\msx@41
\mathchardef\vartriangleright="3\msx@42
\mathchardef\vartriangleleft="3\msx@43
\mathchardef\trianglerighteq="3\msx@44
\mathchardef\trianglelefteq="3\msx@45
\mathchardef\bigstar="0\msx@46
\mathchardef\between="3\msx@47
\mathchardef\blacktriangledown="0\msx@48
\mathchardef\blacktriangleright="3\msx@49
\mathchardef\blacktriangleleft="3\msx@4A
\mathchardef\vartriangle="3\msx@4D
\mathchardef\blacktriangle="0\msx@4E
\mathchardef\triangledown="0\msx@4F
\mathchardef\eqcirc="3\msx@50
\mathchardef\lesseqgtr="3\msx@51
\mathchardef\gtreqless="3\msx@52
\mathchardef\lesseqqgtr="3\msx@53
\mathchardef\gtreqqless="3\msx@54
\mathchardef\Rrightarrow="3\msx@56
\mathchardef\Lleftarrow="3\msx@57
\mathchardef\veebar="2\msx@59
\mathchardef\barwedge="2\msx@5A
\mathchardef\doublebarwedge="2\msx@5B
\mathchardef\angle="0\msx@5C
\mathchardef\measuredangle="0\msx@5D
\mathchardef\sphericalangle="0\msx@5E
\mathchardef\varpropto="3\msx@5F
\mathchardef\smallsmile="3\msx@60
\mathchardef\smallfrown="3\msx@61
\mathchardef\Subset="3\msx@62
\mathchardef\Supset="3\msx@63
\mathchardef\Cup="2\msx@64

\mathchardef\Cap="2\msx@65

\mathchardef\curlywedge="2\msx@66
\mathchardef\curlyvee="2\msx@67
\mathchardef\leftthreetimes="2\msx@68
\mathchardef\rightthreetimes="2\msx@69
\mathchardef\subseteqq="3\msx@6A
\mathchardef\supseteqq="3\msx@6B
\mathchardef\bumpeq="3\msx@6C
\mathchardef\Bumpeq="3\msx@6D
\mathchardef\lll="3\msx@6E

\mathchardef\ggg="3\msx@6F

\mathchardef\circledS="0\msx@73
\mathchardef\pitchfork="3\msx@74
\mathchardef\dotplus="2\msx@75
\mathchardef\backsim="3\msx@76
\mathchardef\backsimeq="3\msx@77
\mathchardef\complement="0\msx@7B
\mathchardef\intercal="2\msx@7C
\mathchardef\circledcirc="2\msx@7D
\mathchardef\circledast="2\msx@7E
\mathchardef\circleddash="2\msx@7F
\def\ulcorner{\delimiter"4\msx@70\msx@70 }
\def\urcorner{\delimiter"5\msx@71\msx@71 }
\def\llcorner{\delimiter"4\msx@78\msx@78 }
\def\lrcorner{\delimiter"5\msx@79\msx@79 }
\def\yen{\mathhexbox\msx@55 }
\def\checkmark{\mathhexbox\msx@58 }
\def\circledR{\mathhexbox\msx@72 }
\def\maltese{\mathhexbox\msx@7A }
\mathchardef\lvertneqq="3\msy@00
\mathchardef\gvertneqq="3\msy@01
\mathchardef\nleq="3\msy@02
\mathchardef\ngeq="3\msy@03
\mathchardef\nless="3\msy@04
\mathchardef\ngtr="3\msy@05
\mathchardef\nprec="3\msy@06
\mathchardef\nsucc="3\msy@07
\mathchardef\lneqq="3\msy@08
\mathchardef\gneqq="3\msy@09
\mathchardef\nleqslant="3\msy@0A
\mathchardef\ngeqslant="3\msy@0B
\mathchardef\lneq="3\msy@0C
\mathchardef\gneq="3\msy@0D
\mathchardef\npreceq="3\msy@0E
\mathchardef\nsucceq="3\msy@0F
\mathchardef\precnsim="3\msy@10
\mathchardef\succnsim="3\msy@11
\mathchardef\lnsim="3\msy@12
\mathchardef\gnsim="3\msy@13
\mathchardef\nleqq="3\msy@14
\mathchardef\ngeqq="3\msy@15
\mathchardef\precneqq="3\msy@16
\mathchardef\succneqq="3\msy@17
\mathchardef\precnapprox="3\msy@18
\mathchardef\succnapprox="3\msy@19
\mathchardef\lnapprox="3\msy@1A
\mathchardef\gnapprox="3\msy@1B
\mathchardef\nsim="3\msy@1C
\mathchardef\napprox="3\msy@1D
\mathchardef\varsubsetneq="3\msy@20
\mathchardef\varsupsetneq="3\msy@21
\mathchardef\nsubseteqq="3\msy@22
\mathchardef\nsupseteqq="3\msy@23
\mathchardef\subsetneqq="3\msy@24
\mathchardef\supsetneqq="3\msy@25
\mathchardef\varsubsetneqq="3\msy@26
\mathchardef\varsupsetneqq="3\msy@27
\mathchardef\subsetneq="3\msy@28
\mathchardef\supsetneq="3\msy@29
\mathchardef\nsubseteq="3\msy@2A
\mathchardef\nsupseteq="3\msy@2B
\mathchardef\nparallel="3\msy@2C
\mathchardef\nmid="3\msy@2D
\mathchardef\nshortmid="3\msy@2E
\mathchardef\nshortparallel="3\msy@2F
\mathchardef\nvdash="3\msy@30
\mathchardef\nVdash="3\msy@31
\mathchardef\nvDash="3\msy@32
\mathchardef\nVDash="3\msy@33
\mathchardef\ntrianglerighteq="3\msy@34
\mathchardef\ntrianglelefteq="3\msy@35
\mathchardef\ntriangleleft="3\msy@36
\mathchardef\ntriangleright="3\msy@37
\mathchardef\nleftarrow="3\msy@38
\mathchardef\nrightarrow="3\msy@39
\mathchardef\nLeftarrow="3\msy@3A
\mathchardef\nRightarrow="3\msy@3B
\mathchardef\nLeftrightarrow="3\msy@3C
\mathchardef\nleftrightarrow="3\msy@3D
\mathchardef\divideontimes="2\msy@3E
\mathchardef\varnothing="0\msy@3F
\mathchardef\nexists="0\msy@40
\mathchardef\mho="0\msy@66
\mathchardef\thorn="0\msy@67
\mathchardef\beth="0\msy@69
\mathchardef\gimel="0\msy@6A
\mathchardef\daleth="0\msy@6B
\mathchardef\lessdot="3\msy@6C
\mathchardef\gtrdot="3\msy@6D
\mathchardef\ltimes="2\msy@6E
\mathchardef\rtimes="2\msy@6F
\mathchardef\shortmid="3\msy@70
\mathchardef\shortparallel="3\msy@71
\mathchardef\smallsetminus="2\msy@72
\mathchardef\thicksim="3\msy@73
\mathchardef\thickapprox="3\msy@74
\mathchardef\approxeq="3\msy@75
\mathchardef\succapprox="3\msy@76
\mathchardef\precapprox="3\msy@77
\mathchardef\curvearrowleft="3\msy@78
\mathchardef\curvearrowright="3\msy@79
\mathchardef\digamma="0\msy@7A
\mathchardef\varkappa="0\msy@7B
\mathchardef\hslash="0\msy@7D
\mathchardef\hbar="0\msy@7E
\mathchardef\backepsilon="3\msy@7F
\def\Bbb{\ifmmode\let\next\Bbb@\else
 \def\next{\errmessage{Use \string\Bbb\space only in math mode}}\fi\next}
\def\Bbb@#1{{\Bbb@@{#1}}}
\def\Bbb@@#1{\fam\msyfam#1}

\catcode`\@=12
\font\teneusmf=eufm10 scaled 1200
\font\seveneusmf=eufm8
\font\fiveeusmf=eufm6
\newfam\eusmffam
\textfont\eusmffam=\teneusmf
\scriptfont\eusmffam=\seveneusmf
\scriptscriptfont\eusmffam=\fiveeusmf
\def\gothi#1{{\fam\eusmffam\relax#1}}
\font\teneusm=eusm10 scaled 1200
\font\seveneusm=eusm8
\font\fiveeusm=eusm6
\newfam\eusmfam
\textfont\eusmfam=\teneusm
\scriptfont\eusmfam=\seveneusm
\scriptscriptfont\eusmfam=\fiveeusm

\font\teneusmc=cmsy10 scaled 1200
\font\seveneusmc=cmsy8
\font\fiveeusmc=cmsy6
\newfam\eusmcfam
\textfont\eusmcfam=\teneusmc
\scriptfont\eusmcfam=\seveneusmc
\scriptscriptfont\eusmcfam=\fiveeusmc
\def\cally#1{{\fam\eusmcfam\relax#1}}

\newlength{\dinwidth}
\newlength{\dinmargin}
\newlength{\sectionnumbersize}
\newlength{\sectionsize}
\newcommand{\dgsection}[1]{
\refstepcounter{section}
\bigskip
\noindent
{\Large\bf
\makebox[\sectionnumbersize][l]{\arabic{section}}
\begin{minipage}[t]{\sectionsize}
#1
\dgsectionbreak
\vspace{-0.1cm}
\end{minipage}
}
\vspace{-0.7cm}
\nopagebreak

}

\newcommand{\dgsectionbreak}{\\*[-0.2cm]}

\begin{document}

\begin{titlepage}

\renewcommand{\thefootnote}{\fnsymbol{footnote}}
\setcounter{footnote}{0}

\hfill
\hspace*{\fill}
\begin{minipage}[t]{5cm}
CERN-TH.\cernnr
\end{minipage}


\vspace{0.5cm}
\begin{center}

{\LARGE\bf On the Space-Time Geometry\\ of Quantum Systems}

\vspace{0.5cm}


\vspace{0.1cm}

{\bf Dirk Graudenz}$\;${\it\footnote[1]{{\em Electronic
mail addresses: graudenz \char64{ }cernvm.cern.ch,
i02gau \char64{ }dsyibm.desy.de}} \\
\vspace{0.1cm}
Theoretical Physics Division, CERN\\
CH--1211 Geneva 23\\
}

\end{center}

\begin{abstract}
We describe the time evolution of quantum systems in a classical background
space-time by means of a covariant derivative in an infinite dimensional
vector bundle. The corresponding parallel transport operator along a timelike
curve $\cC$ is interpreted as the time evolution operator of an observer
moving along $\cC$. The holonomy group of the connection, which can be
viewed as a group of local symmetry transformations,  and the set of
observables have to satisfy certain consistency conditions.
Two examples related to local $\mbox{SO}(3)$ and $\mbox{U}(1)$-symmetries,
respectively, are discussed in detail. The theory developed in this paper may
also be useful to analyze situations where the underlying space-time manifold
has closed timelike curves.
\end{abstract}

\renewcommand{\thefootnote}{\arabic{footnote}}
\setcounter{footnote}{0}

\vfill
\noindent
\begin{minipage}[t]{5cm}
CERN-TH.\cernnr\\
\vdate
\end{minipage}
\vspace{1cm}
\end{titlepage}


\newpage

\begin{figure}[htb]
\centerline{\epsfig
{figure=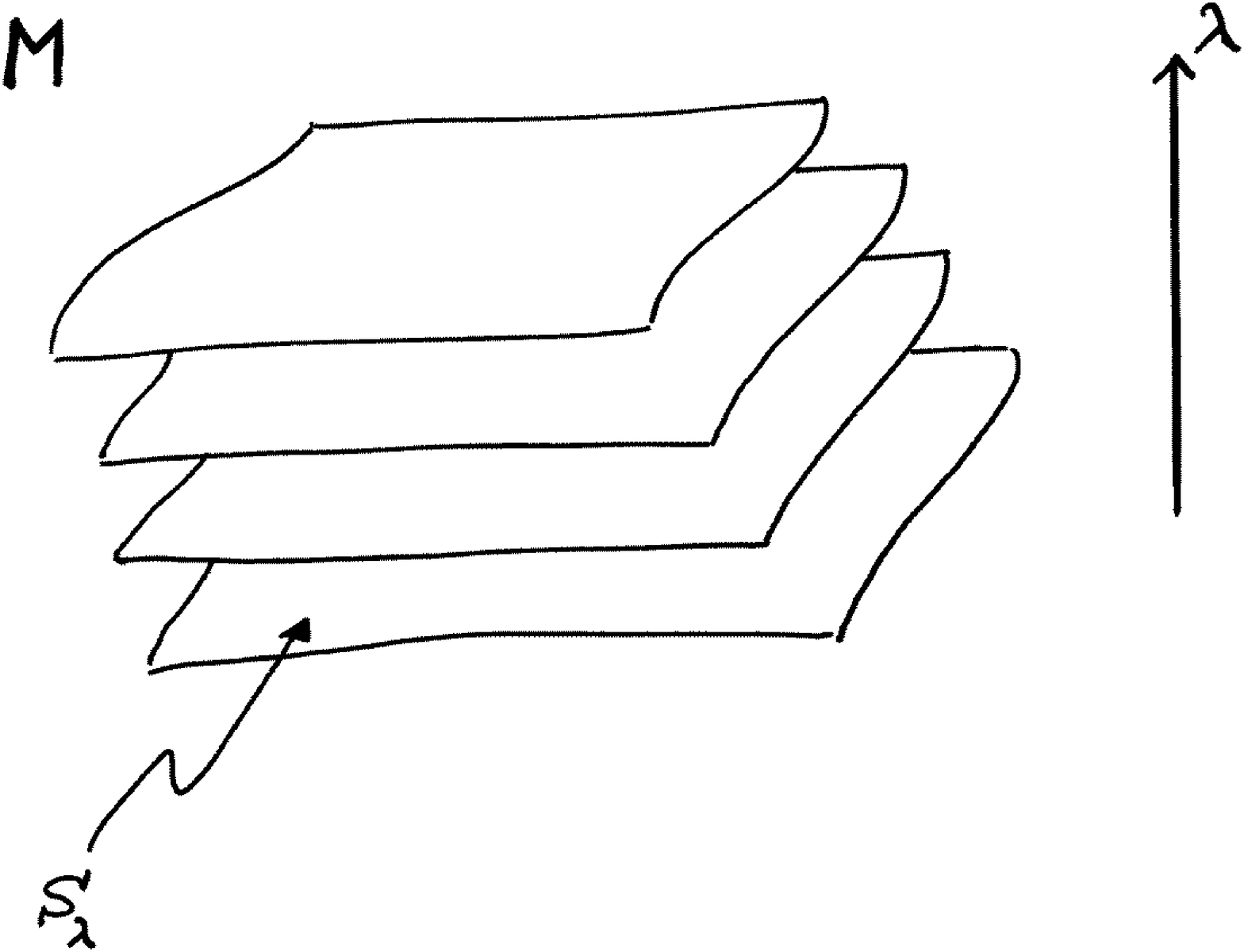,width=10cm,clip=}
           }
{\em \caption{\label{foliation} Foliation of space-time by
spacelike hypersurfaces.}}
\end{figure}

\dgsection{Introduction}
\label{intro}
\noindent
The state vector in quantum theory is an object that describes the knowledge
about a physical system. In the case of a local quantum field theory in a
possibly curved space-time background, the maximum knowledge about the system
can be encoded in a state vector $\Psi_S$ that is defined for a spacelike
hypersurface $S$ \cite{1}.
This is the case because observables whose space-time supports are separated
by a spacelike distance commute and can thus be separately diagonalized.
In the definition of states $\Psi_S$
it is implicitly assumed that the space-time manifold
$M$ can be foliated by a set of spacelike hypersurfaces $S_\lambda$,
where $\lambda$ is a real parameter;
see Fig.~\ref{foliation}. The evolution of the system is then given
by a Schr\"odinger
equation\footnote{For simplicity we assume $H_\lambda$ to be
anti-Hermitian in order to avoid factors of $i$.} of the form
\beq
\frac{\di}{\di\lambda}\,\Psi_{S_\lambda}\,=\,H_\lambda\,\Psi_{S_\lambda}.
\eeq
$\lambda$ plays the r\^{o}le of a global time coordinate.
The approach of slicing space-time by spacelike hypersurfaces
has the disadvantage that
the time evolution is not easily generalized to space-time manifolds
with closed timelike curves or similar causal obstructions,
since sometimes it is not possible to find a global foliation $S_\lambda$.

Here we try to give a formulation of quantum theory that in principle allows a
description of the time evolution of
a system even in the problematic cases just mentioned.
The key observation is that knowledge is always knowledge by an
observer $B$. Let $\cC_\tau$ be $B$'s worldline, where $\tau$
parametrizes $\cC$; one can think of $\tau$ as $B$'s eigentime.
We will associate with every curve $\cC$ a state vector $\Psi_\cC(\tau)$
depending on the curve $\cC$ and on the curve parameter $\tau$.
For $B$, the state evolution will be described by a Hamiltonian
operator $H_\cC(\tau)$ in the form of a Schr\"{o}dinger equation
\beq
\label{S1}
\frac{\di}{\di\tau}\,\Psi_\cC(\tau)\,=\,H_\cC(\tau)\,\Psi_\cC(\tau).
\eeq
The Hamiltonian operator depends on the space-time curve $\cC$.
By integrating Eq.~(\ref{S1}) we may define a unitary evolution operator
$U_\cC(\tau,\tau_0)$ satisfying the equation
\beq
\label{S2}
\partial_\tau\,U_\cC(\tau,\tau_0)\,
=\,H_\cC(\tau)\,U_\cC(\tau,\tau_0)
\eeq
with the initial condition $U_\cC(\tau_0,\tau_0)=\opone$.
For a curve $\cD:[a,b]\rightarrow M$ we may define
$U_\cD$ by $U_\cD(b,a)$ (i.e. by the evolution operator from
the beginning to the end of the curve).
We will assume, for simplicity,
that all observables under consideration are scalar quantities
such that two observers $B_1$, $B_2$ at the same space-time point $x$
could use the same state vector $\Psi$ regardless of the relative orientation
of their coordinate systems and their relative velocity.

\begin{figure}[htb]
\centerline{\epsfig
{figure=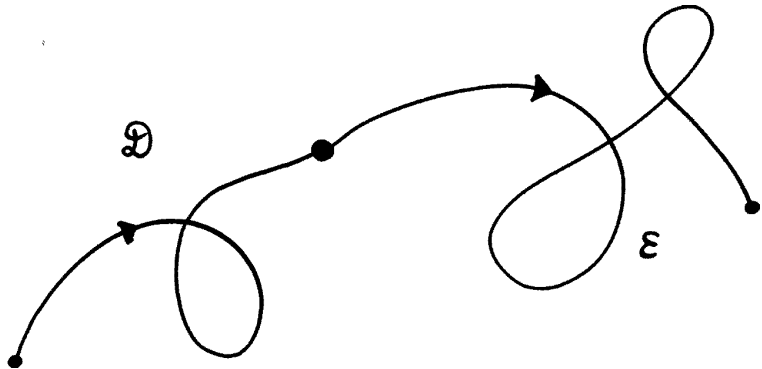,width=10cm,clip=}
           }
{\em \caption{\label{fig1} Concatenation of two curves.}}
\end{figure}

For two curves $\cD$, $\cE$ with the property that the endpoint of $\cD$
is the starting point of $\cE$, let $\cE\circ\cD$ be the concatenation
of $\cD$ and $\cE$, i.e. the curve that is obtained by first following
$\cD$ and then $\cE$; see Fig.~\ref{fig1}.

We assume that the specific parametrisations of the curves are irrelevant;
physics should be independent of the coordinates chosen.
We certainly require that the evolution operators satisfy
\beq
\label{S3}
U_{\cE\circ\cD}\,=\,U_\cE\,\circ\,U_\cD,
\eeq
which is a compatibility condition with Eq.~(\ref{S2}).
Moreover, we require that
\beq
\label{S4}
U_{\cD^{-1}}\,=\,U_\cD^{-1},
\eeq
where $\cD^{-1}$ is the curve $\cD$ traversed in the opposite direction.
All these properties are precisely those of parallel transport operators.
This fact suggests a formulation in the language of differential geometry,
which will be developed in Section~\ref{diffg}.

The connection of the vector bundle introduced there will, in general,
possess a non-trivial holonomy group. Since the expectation values,
predicted by two observers
$B,\,C$ for a measurement at a
system should not differ, the non-trivial
holonomy leads to certain consistency conditions.
It turns out that the consistency conditions can be interpreted as
a structure of local symmetry transformations. The group of these
symmetry transformations at a space-time point $x$
is a representation of the loop group at $x$.
We investigate this structure in Section~\ref{consistency}.
Two examples related to local
$\mbox{SO}(3)$ and $\mbox{U}(1)$-symmetries,
respectively, are discussed in Section~\ref{applications}.
The paper closes with a summary and some conclusions.

\dgsection{A Formulation in the Language of Differential\dgsectionbreak
Geometry}
\label{diffg}
\noindent
We formulate in this section the evolution of quantum systems
in a differential geometric
framework\footnote{We do not attempt to formulate everything in a
mathematically rigorous way and rather concentrate on the
conceptual development instead.
\newcounter{riglabel}
\setcounter{riglabel}{\value{footnote}}
}. As we have seen in the preceding
section, the evolution operators $U_\cD$ have the properties of
parallel transport operators. In general, they act on a fibre
bundle $\pi_G:G\rightarrow M$ over the
(connected) space-time manifold $M$.
The fibre $G_x:=\pi_G^{-1}(x)$ over $x$ is isomorphic to a Hilbert
space $\cH$.
The state of the quantum system, as described by an observer at
$x\in M$, is an element of $G_x$.

For a curve $\cD:[a,b]\rightarrow M$,
$U_\cD$ maps $G_{\cD_a}$ onto $G_{\cD_b}$.
Under suitable differentiablity conditions the parallel transport operators
$U_\cD$ have an infinitesimal representation in terms of a connection
or covariant derivative $D$ on the bundle $\pi_G$.
In a local coordinate system $\Phi$ given by
\beq
G\Big|_W\,\stackrel{\Phi}{\rightarrow}\,W\,\times \cH,\quad W\subset M,
\quad G\Big|_W\,\defeq\,\bigcup\limits_{x\in W}\,G_x,
\eeq
$D$ can be expressed as $D=\di-H$, where
$\di=\partial_\mu\cdot \di x^\mu$ is the differential
on $W$, and $H=H_\mu\,\di x^\mu$ is an operator-valued one-form on $M$.
As will be seen shortly, $H$ transforms like a gauge field under
a change of the frame.
In the local frame $\Phi$, the evolution operator $H_\cC(\tau)$
introduced above is $H_\cC(\tau)\,=\,H_\mu(\cC_\tau)\,\dot{\cC}^\mu_\tau$,
where the dot denotes differentiation with respect to the curve parameter.
Therefore,
\beq
\label{S6}
\partial_\tau\,U_\cC(\tau,\tau_0)\,
=\,H_\mu(\cC_\tau)\,\dot{\cC}^\mu_\tau\,U_\cC(\tau,\tau_0).
\eeq

Under a change of coordinates
\beq
\label{chco}
\chi=\Phi_2\circ \Phi_1^{-1},
\eeq
the representative $(x,\Psi)=\Phi_1(\psi)$ of a vector $\psi$
over $x$ in the frame $\Phi_1$
is mapped via
$\chi$ into the representative
$(x,V(x)\Psi)=\Phi_2(\psi)$
in the frame $\Phi_2$ , where $V(x)$ is a unitary operator on $\cH$
(we assume that the structure group of $\pi_G$ is
a group of unitary operators).
If $H_{\Phi_1}$ is the
generator of parallel translations in the frame $\Phi_1$,
then the corresponding generator $H_{\Phi_2}$ in the frame $\Phi_2$
is
\beq
H_{\Phi_2}(x)\,=\,V(x)\,H_{\Phi_1}(x)\,V(x)^{-1}\,+\,\di V(x)\,V(x)^{-1},
\eeq
satisfying
$(\di-H_{\Phi_2})(V\Psi)=V(\di-H_{\Phi_1})\Psi$.
This transformation law is that of a gauge field. The inhomogeneous
term arises because the unitary transformation operator $V$
will, in general, depend on the space-time point $x$.

Up to now, we have just considered the evolution of the state vectors.
We now discuss the question of observables.
Let $\gM$ be a set of observables.
We think of $A\in \gM$ as some scalar observable like the field
strength of a scalar field. In general, however, $A$ may be any
scalar quantity that can be measured by means of a suitable
`infinitesimal measuring device' at any space-time point.
It should be noted that the observables $A$ do not carry a space-time index.
In order to associate observables to space-time points, we assume that
there is a mapping
$\varphi$ of the set of observables into the algebra of operators
on $G$, such that $\varphi(A)\equiv A_\varphi$ for
$A \in \gM$ maps every fibre $G_x$ into itself.

In order to calculate expectation values of observables, a Hermitian
inner product $P(\psi,\xi)\equiv \langle\psi|\xi\rangle$ for vectors
$\psi,\,\xi$ from the same fibre of $G$ is needed.
The quantity
$\langle\psi|A_\varphi|\psi\rangle
/\langle\psi|\psi\rangle$
is interpreted as the expectation value for the
observable $A$ at the space-time point $x$ if the state of the system
is described by an observer $B$ at $x$ by means of the state
$\psi\in G_x$.

Let the prototype Hilbert space $\cH$ be equipped with the Hermitian
inner product $[\cdot|\cdot]$. Then the expression
$\langle\psi|A_\varphi|\xi\rangle$  for
$\psi,\,\xi\in G_x$ can be written in a local frame $\Phi$ as
\beq
\langle\psi|A_\varphi|\xi\rangle =
[\Phi(\psi)|P_\Phi(x)A_{\varphi\Phi}(x)|\Phi(\xi)],
\eeq
where $P_\Phi(x)$ is a Hermitian operator (Hermitian with respect to
$[\cdot|\cdot]$) representing the inner product $P$ in the
fibre $G_x$ over $x$, and
$A_{\varphi\Phi}(x)$ is a Hermitian operator (Hermitian with respect to
$[\cdot|P_\Phi(x)\cdot]$) corresponding to the observable $A$.
Under the change of frame in Eq.~(\ref{chco})
the inner product $P$ and the operators $A_\varphi$ transform like
\beqn
P_{\Phi_2}(x) &=& V(x)P_{\Phi_1}(x)V(x)^{-1},\\
A_{\Phi_2}(x) &=& V(x)A_{\Phi_1}(x)V(x)^{-1}.
\eeqn
We assume that the inner product $P$ and the
covariant derivative are compatible. In terms of the parallel transport
operators this means that
\beq
\langle U_\cD\psi|U_\cD\xi\rangle=\langle\psi|\xi\rangle,
\eeq
where $\psi,\,\xi$ are vectors from the fibre of $G$ over the starting point of
$\cD$.

So far we have considered the prediction by an observer $B$ at $x$
of expectation
values for measurements of observables at $x$.
We now give a prescription for the prediction of expectation values
of measurements at $y\neq x$. The idea is simple: let $\psi_B$ be the state
vector used by $B$ at $x$. Choose a curve $\cD:[a,b]\rightarrow M$
connecting $x=\cD_a$ and $y=\cD_b$ and transport $\psi_B$ from $x$ to $y$
via $\psi_B^\cD= U_\cD\psi_B$ along $\cD$.
Define the expectation value for a measurement of $A$ at $y$ to be
$\langle\psi_B^\cD|A_\varphi(y)|\psi_B^\cD\rangle$
(we assume $\langle\psi_B^\cD|\psi_B^\cD\rangle=1$).
In general, $\psi_B^\cD$ depends on the curve $\cD$ chosen. The
consistency conditions resulting from this dependence will be discussed
in the next section.
As a preparation of this discussion, we consider the curvature $F=D^2$
of the connection $D$. $F$ is an operator-valued two-form, mapping
two tangent vectors $v,\,w$ of the tangent space $T_xM$ of $M$ at $x$
and a vector $\psi\in G_x$ into a vector $F(v,w)\psi\in G_x$.
$F$ is a measure for how much the parallel transport
around an infinitesimal closed loop fails to be the identity,
\beq
\label{fieldstrength}
U_\beta=\opone+\tau^2F(v,w),
\eeq
where $\beta$ is the infinitesimal closed loop ($\tau\ll 1$)
connecting the points $(x,x+\tau v,x+\tau v+\tau w,x+\tau w,x)$.

\begin{figure}[htb]
\centerline{\epsfig
{figure=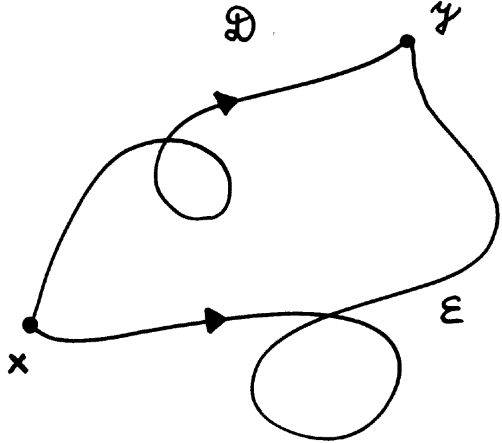,width=7cm,clip=}
           }
{\em \caption{\label{figjoin} Path independence of expectation values.}}
\end{figure}

\dgsection{Consistency Conditions and the Local Symmetry
\dgsectionbreak
Structure}
\label{consistency}
\noindent
We now consider the dependence
on the curve $\cD$ joining $x$ and $y$
of predictions for measurements at $y$ by an
observer $B$ at $x$.
Let $\cE$ be another curve joining $x$ and $y$;
see Fig.~\ref{figjoin}.
The prediction will then be
$\langle\psi_B^\cE|A_\varphi(y)|\psi_B^\cE\rangle$.
Certainly the expectation value should not depend
on the path chosen, so we require
\beq
\langle\psi_B^\cD|A_\varphi(y)|\psi_B^\cD\rangle=
\langle\psi_B^\cE|A_\varphi(y)|\psi_B^\cE\rangle.
\eeq
Thus,
\beq
\langle U_\cD\psi_B|A_\varphi(y)|U_\cD\psi_B\rangle=
\langle U_\cE\psi_B|A_\varphi(y)|U_\cE\psi_B\rangle.
\eeq
This condition can be rewritten as
\beq
\langle U_\cD\psi_B|U_\alpha A_\varphi(y)
U_\alpha^{-1} - A_\varphi(y)|U_\cD\psi_B\rangle=0.
\eeq
Here the closed loop $\alpha$ is defined by $\alpha=\cD\circ\cE^{-1}$.
The condition has to be satisfied for all $y$, for all
$\psi_B\in G_x$ and for all curves $\cD,\,\cE$ joining $x$ and $y$.
Therefore, since $U_\cD$ is an isomorphism,
the condition
\beq
\langle \xi|U_\alpha A_\varphi(y)
U_\alpha^{-1} - A_\varphi(y)|\xi\rangle=0
\eeq
has to hold for all
$\xi\in G_y$ and for all closed loops $\alpha$
located at $y$. As a consequence,
\beq
\label{commute}
[U_\alpha,A_\varphi(y)]=0
\eeq
if all vectors $\xi$ are physical.

The set of all parallel transport operators $U_\alpha$
along closed loops $\alpha$ located at the same point $y$
forms a group, the so-called holonomy group $\gH(y)$ of the connection
$D$ at $y$. The relation Eq.~(\ref{commute})
expresses the fact that the holonomy
group at $y$ has to commute with all observables at $y$.
The $U_\alpha$ transform states into states and have to leave observables
invariant. This is to say that they act as local symmetry transformations.
These transformations are {\em not} related to the transformations
of the change of coordinates Eq.~(\ref{chco}) in Section~\ref{diffg}.
On the contrary, they {\em do} transform state vectors non-trivially.

\begin{figure}[htb]
\centerline{\epsfig
{figure=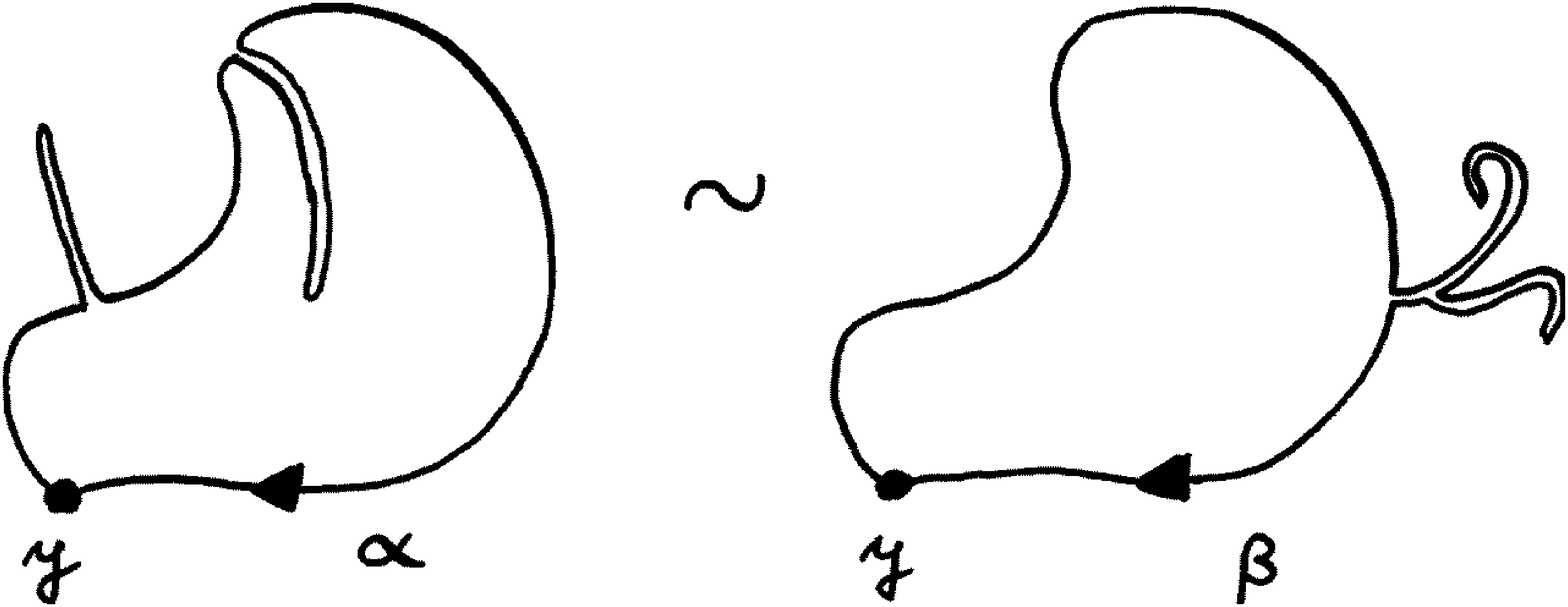,width=10cm,clip=}
           }
{\em \caption{\label{equiv} Equivalent closed loops.}}
\end{figure}

The holonomy group $\gH(y)$ is a representation of the loop group $\gL(y)$
of $M$ at $y$
\cite{2}. The loop group as a set consists of classes $[\alpha]$
of closed loops at $y$, where two loops $\alpha,\,\beta$ are equivalent,
$\alpha\sim\beta$, if $\beta$ can be obtained from $\alpha$
by inserting or deleting curve segments of the form
$\cF\circ\cF^{-1}$, see Fig.~\ref{equiv}. Group multiplication is defined by
$[\alpha]\circ[\beta]=[\alpha\circ\beta]$, and the inverse by
$[\alpha]^{-1}=[\alpha^{-1}]$. Let $\gL^0(y)$ be the subgroup
of $\gL(y)$ consisting of the classes $[\alpha]$ of closed loops $\alpha$
which are homotopic to $0$, i.e. which can be smoothly contracted
to the constant curve at $y$.
Let $\gH^0(y)$ be the restricted holonomy group consisting of operators
$U_\alpha$ for $[\alpha]\in\gL^0(y)$ (see \cite{3}).
It is known that $\gH^0(y)$ is a
normal subgroup of $\gH(y)$, and that the factor
group $\gH(y)/\gH^0(y)$ is discrete\footnote{This is true in the case of
a finite dimensional bundle. We assume that the same is true
in the infinite dimensional
case$^{\mbox{\scriptsize\arabic{riglabel}}}$.
}.
The first homotopy
group $\pi_1(M,y)$ of the space-time manifold $M$
has a representation in this discrete group.
Thus the topology of space-time
is partly reflected in the structure of the observable algebra in the
form of a discrete factor group of the group of symmetry transformations.
The condition $[U_\alpha,A_\varphi(y)]=0$ for
$\{\alpha\}\in \pi_1(M,y)$ ($\{\alpha\}$ being the homotopy class
of $\alpha$) can alternatively be interpreted as a discrete
`quantization condition'.

If the set of observables $\gM$ is complete in the sense that
the condition $\mbox{tr}(A_\varphi^\cD\rho)=0$
for a density matrix $\rho$, for all curves $\cD$
and for all observables
$A\in\gM$ has the consequence that $\rho=0$, then
for all closed loops $\alpha$ the corresponding parallel transport operator
is a phase transformation, $U_\alpha=e^{i\varphi_\alpha}$.
We can construct a `projective bundle'
$\pi_{\sP G}:\sP G\rightarrow M$ by identifying
all $\psi,\psi^\prime\in G_x\setminus\{0\}$
with $\psi=\lambda\psi^\prime$,
$\lambda\in\sC\setminus\{0\}$. The $U_\cD$ define operators
$\sP U_\cD$ acting as parallel transport operators in $\pi_{\sP G}$.
In the case just mentioned,
$\sP U_\cD=\mbox{id}$. Therefore the bundle $\pi_{\sP G}$ is
trivial since parallel transport does not depend on the path chosen,
and thus a global frame can be constructed.

\begin{figure}[htb]
\centerline{\epsfig
{figure=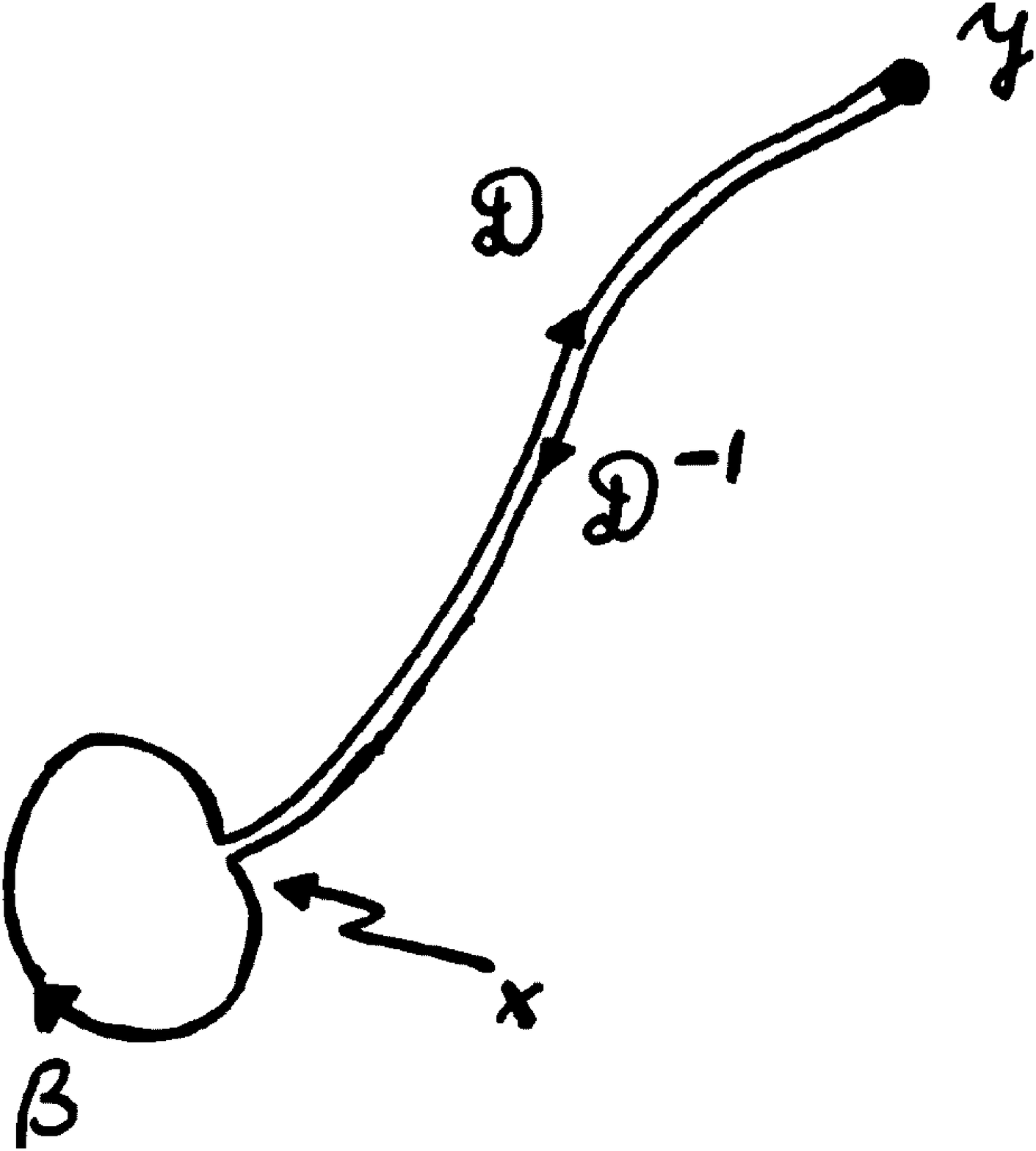,width=5cm,clip=}
           }
{\em \caption{\label{smallcurve} The local consistency condition.}}
\end{figure}

The condition $[U_\alpha,A_\varphi(y)]=0$ for all null homotopic loops
can be rewritten in terms of the curvature $F$. For a curve $\cD$ joining
$x$ and $y$ we define the operator $A_\varphi^\cD$
on $G_x$ by the expression
$U_\cD^{-1}A_\varphi(y)U_\cD$.
$A_\varphi^\cD$ maps $G_x$ into $G_x$ and obviously fulfils
the relation
\beq
\langle\psi_B^\cD|A_\varphi(y)|\psi_B^\cD\rangle=
\langle\psi_B|A_\varphi^\cD|\psi_B\rangle.
\eeq
Now we consider a closed loop $\beta$ at $x$.
$\alpha=\cD\circ\beta\circ\cD^{-1}$ is a closed loop at $y$, see
Fig.~\ref{smallcurve}. The consistency condition
$[U_\alpha,A_\varphi(y)]=0$ can be rewritten as
$[U_\beta,A_\varphi^\cD]=0$. If we now consider the case of an infinitesimal
closed loop $\beta$ at $x$, then the condition finally reads
$[F_{\mu\nu}(x),A_\varphi^{\cD}]=0$, cf. Eq.~(\ref{fieldstrength}).
The curvature tensor at $x$ has to commute with all observables
parallel-transported to $x$ along arbitrary curves.

\begin{figure}[htb]
\centerline{\epsfig
{figure=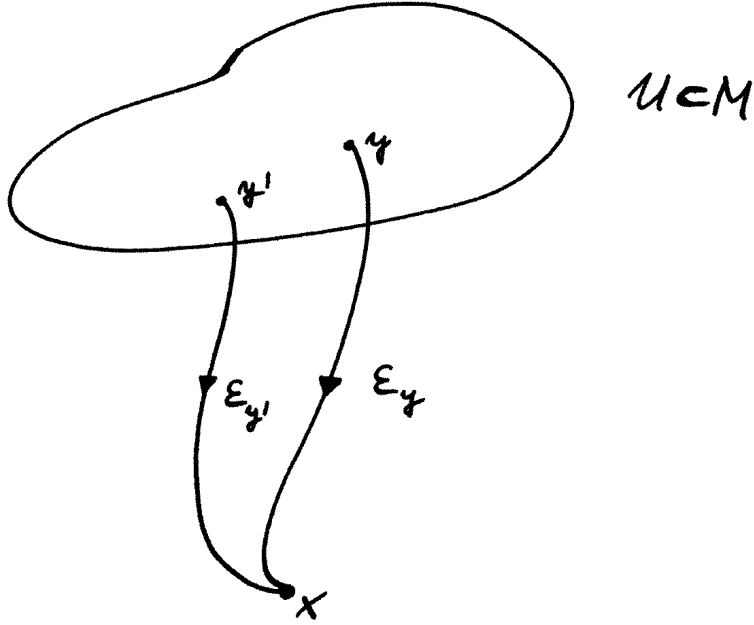,width=10cm,clip=}
           }
{\em \caption{\label{contraction} A contraction of $U$ to $x$.}}
\end{figure}

Assume that an observer $B$ at $x\in M$ wants to give a description of
measurements performed in a remote space-time region $U\subset M$.
We assume that $U$ is contractible. Then there is a map
\beq
h\,:\,U\times I\,\rightarrow\,M,
\eeq
where $I=[0,1]$, with the property that $h(y,0)=y$, $h(y,1)=x$
for all $y\in U$ (see Fig.~\ref{contraction}).
Mathematically, $h$ is a homotopy of the inclusion map $\iota_U:
U\rightarrow M$ and the constant map $c_x:U\rightarrow M$,
$c_x(y)=x$.

Define the family of curves $\cE_y$ by
$\cE_y(\tau)=h(y,\tau)$. Thus $\cE_y$ joins $y$ and $x$.
The curves $\cE_y$ allow us to define a local representation
of $U$ at $x$ as follows.
Define operators
\beq
A_{\varphi h}(y)\,\defeq \,A_{\varphi\cE_y}\,\defeq\,
U_{\cE_y}\,A_\varphi(y)\,U_{\cE_y}^{-1}\,=\,A_\varphi^{\cE_y^{-1}}.
\eeq
$A_{\varphi h}(y)$ is an operator on the fibre $G_x$.
The expectation value for $A$ at $y\in U$, given that $B$'s state vector is
$\psi_B$, is
$\langle\psi_B|A_{\varphi h}(y)|\psi_B\rangle$.
In a local frame, the operators $A_{\varphi \cE_y|_\tau}$
satisfy the equation
\beq
\partial_\tau A_{\varphi \cE_y|_\tau}
=[H_\mu(\cE_\tau)\dot{\cE}_\tau^\mu,A_{\varphi \cE_y|_\tau}].
\eeq
Here $\cD|_\tau$ is the restriction of the curve
$\cD:[a,b]\rightarrow M$ to the interval
$[a,\tau]$.

\begin{figure}[htb]
\centerline{\epsfig
{figure=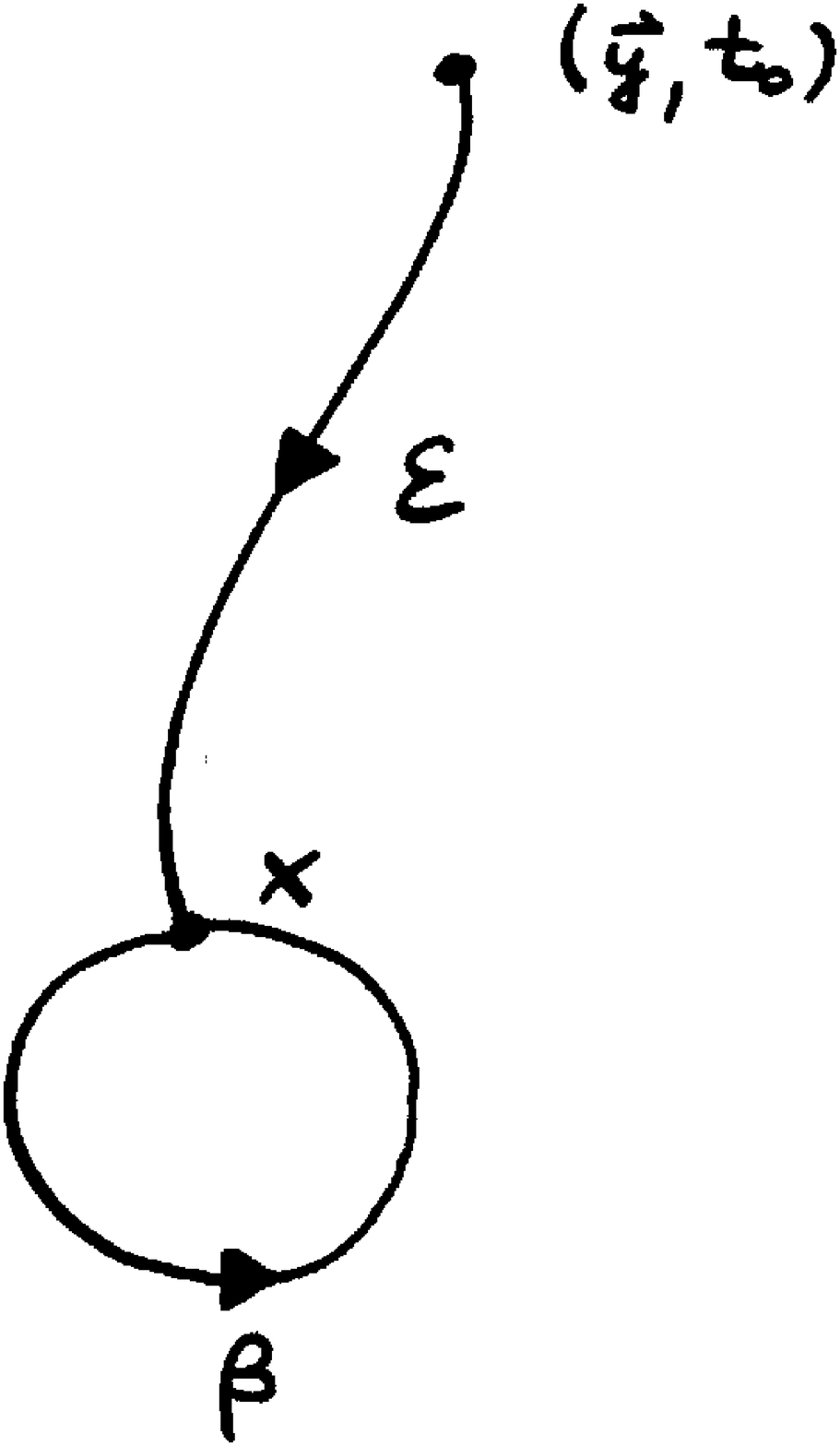,width=4cm,clip=}
           }
{\em \caption{\label{gaugeloop} A local symmetry transformation of
a field at $(\vec{y},t_0)$.}}
\end{figure}

\dgsection{Applications}
\label{applications}
\noindent
In order to develop instructive examples, we
study a simple class of systems. We assume that the space-time manifold
is a product $M=S\times T$ of a `space' manifold $S$ and a
one-dimensional `time' manifold
$T$. The bundle under consideration is a trivial bundle
$G=M\times\cH$. A point in $G$ is of the form
$(\vec{x},t,\psi)$, where $\vec{x}\in S$, $t \in T$,
$\psi \in \cH$. Let $t_0\in T$ be a fixed instant of time.
Define $S_t=S\times\{t\}$, and in particular $S_0=S_{t_0}$.
We assume that there are fields $F_i(\vec{x})$ defined on
$S_0$ fulfilling
\beq
[F_i(\vec{x}),F_i(\vec{y})]=0
\eeq
for $\vec{x}\neq\vec{y}$.
$i$ is an index labelling different fields.
The parallel transport generators $H_\mu$ are assumed to have the
following property:
\beq
\label{star1}
[H_\mu(\vec{x},t),F_i(\vec{y})]=\lambda_{i\mu}^j(\vec{x},t,\vec{y})
F_j(\vec{y}),
\eeq
where the $\lambda_{i\mu}^j(\vec{x},t,\vec{y})$ are complex numbers.
Now consider the one-parameter set of operators
$F_{i\cE|_\tau}=U_{\cE|_\tau}F_i(\vec{y})U_{\cE|_\tau}^{-1}$, where
$\cE$ is a curve with starting point $(\vec{y},t_0)$.
$F_{i\cE|_\tau}$ is a solution of the equation
\beq
\label{star2}
\partial_\tau F_{i\cE|_\tau}
=[H_\mu(\cE_\tau)\dot{\cE}_\tau^\mu,F_{i\cE|_\tau}].
\eeq
We assume $\cE$ to be fixed. In order to solve this equation
explicitly we make the
{\em ansatz}
\beq
\label{star3}
F_{i\cE|_\tau}\,=\,g_i^j(\tau) \,F_j(\vec{y}),\quad
g_i^j(0)=\delta_i^j.
\eeq
Inserting the {\em ansatz}
into Eq.~(\ref{star2}) and making use of Eq.~(\ref{star1}),
we obtain
\beq
\partial_\tau g_i^j(\tau)\,=\,g_i^l(\tau)
\,\lambda_{l\mu}^j(\cE_\tau,\vec{y})\,\dot{\cE}_\tau^\mu.
\eeq
This is an ordinary differential equation for the matrices
$g_i^j(\tau)$. The solution will be denoted by
$g_{\cE|_\tau i}^j$.
The solution for $F_{i\cE|_\tau}$ is then
$F_{i\cE|_\tau}=g_{\cE|_\tau i}^jF_j(\vec{y})$.
In particular, $F_{i\cE}=g_{\cE i}^jF_j(\vec{y})$.
Now we consider a closed loop $\beta$ at $x$, where $\cE$ is assumed to
connect $(\vec{y},t)$ and $x$, see Fig.~\ref{gaugeloop}.

Then, in an obvious notation,
\beq
(F_{i\cE})_\beta\,=\,F_{i\beta\circ\cE}\,=\,g_{\beta\circ\cE i}^j(\vec{y})\,
F_j(\vec{y}),
\eeq
where $g_{\beta\circ\cE i}^j(\vec{y})
=g_{\beta i}^l(\vec{y})g_{\cE l}^j(\vec{y})$,
and
$g_{\cD i}^j(\vec{y})$
for a curve $\cD:[a,b]\rightarrow M$
satisfies the differential equation
\beq
\label{gdiff}
\partial_\tau g_{\cD|_\tau i}^j(\vec{y})\,=\,g_{\cD|_\tau i}^l(\vec{y})
\,\lambda_{l\mu}^j(\cD_\tau,\vec{y})\,\dot{\cD}_\tau^\mu
\eeq
with the initial condition
$g_{\cD|_a i}^j(\vec{y})=\delta_i^j$.

Under parallel transport along $\beta$, $F_{i\cE}$ is transformed into
$g_{\beta i}^j(\vec{y})F_{j \cE}$.
Now assume a homotopy $h$ mapping
$U\subset S_0$ onto $x$. The representation of the fields
$F_i$ at $y=(\vec{y},t_0)$ is
$F_{i\cE_y}$.
The transformation of $F_{i\cE_y}$ depends on
$\vec{y}$, namely
$(F_{i \cE_y})_\beta=g_{\beta i}^j(\vec{y})F_{j \cE_y}$.
Because of this $\vec{y}$-dependence, the transformations
are different for every space-time point: they are local.

It can easily be seen that the condition for the matrices $g_{\cE i}^j$
to be unitary is that
\beq
\label{unitaryg}
\lambda_{j\mu}^i=-\overline{\lambda_{i\mu}^j}.
\eeq
If this is the case, an observable $A$
invariant under holonomy transformations can be constructed by
$A=\sum_i F_i^\dagger F_i^{}$.
Since $g_\beta^\dagger g_\beta^{}=\opone$, we obviously have
$A_\beta=A$.

We now consider two explicit examples.
The first one is based on three fields $T_i$, $i=1,2,3$ on
$M=\sR^3\times\sR$.
We assume that the $T_i$ fulfil the algebra
\beq
[T_i(\vec{x}),T_j(\vec{y})]\,=\,i\,{\epsilon_{ij}}^k\,T_k(\vec{x})\,
\delta(\vec{x}-\vec{y}).
\eeq
The antisymmetric symbol $\epsilon$ is defined by ${\epsilon_{12}}^3=(-1)$,
$\epsilon_{123}=(+1)$.
We define the operators $H_\mu$ by
\beq
H_\mu(x)\,=\,-i\,\int_{S_0}\di \vec{z}\,h_\mu^k(x,\vec{z})
\,T_k(\vec{z}),
\eeq
where the $h_\mu^k(x,\vec{z})$ are real numbers.
The commutators of the $H_\mu$ with the fields are
\beq
[H_\mu(x),T_i(\vec{y})]\,=\,h_\mu^k(x,\vec{y})\,
{\epsilon_{ki}}^j\,T_j(\vec{y}).
\eeq
Therefore in this case
\beq
\lambda_{i\mu}^j(x,\vec{y})\,=\,h_\mu^k(x,\vec{y})\,{\epsilon_{ki}}^j.
\eeq
The condition in Eq.~(\ref{unitaryg}) is obviously satisfied.
The solutions $g_{\cD|_\tau i}^j(\vec{y})$ of Eq.~(\ref{gdiff})
are therefore $\mbox{SO}(3)$-matrices.
For closed curves $\beta$, the parallel transport
operator $U_\beta$ transforms $T_{i\cE_y}$ into the
$\mbox{SO}(3)$-rotated field
$g_{\beta i}^j(\vec{y})T_{j\cE_y}$.
If we moreover assume that the $h_\mu^k(x,\vec{z})$ are
$\vec{z}$-independent, i.e.
\beq
H_\mu(x)=-ih_\mu^k(x)\int_{S_0}\di \vec{z}\,T_k(\vec{z}),
\eeq
then
the rotation matrices $g_{\beta i}^j(\vec{y})$ are
$\vec{y}$-independent; the symmetry is global.

In the second example we assume that the space-time manifold
is a torus $M=\sR^{d-1}\times S^1$; this manifold possesses closed timelike
curves.
We assume fields $q_i(\vec{x})$, $i=1,2$, fulfilling
\beq
[q_i(\vec{x}),q_j(\vec{y})]=0
\eeq
and assume that there are conjugated momenta $p_i(\vec{x})$
such that
\beq
[p_i(\vec{x}),q_j(\vec{y})]=-i\,\delta_{ij}\,\delta(\vec{x}-\vec{y}).
\eeq
The generators $H_\mu$ are defined by
\beq
H_\mu(x)=\frac{1}{2}\,i\,\left[\,\int_{S_0}\di\vec{z}\,
h_\mu(x,\vec{z})
\left(
p_1(\vec{z})q_2(\vec{z})-p_2(\vec{z})q_1(\vec{z})
\right)\,+\,\mbox{h.c.}\right].
\eeq
They satisfy the equation
\beq
[H_\mu(x),q_i(\vec{y})]\,=\,h_\mu(x,\vec{y})\,{\epsilon_i}^j\,q_j(\vec{y}).
\eeq
$\epsilon$ is the antisymmetric symbol with ${\epsilon_1}^2=1$.
Here $\lambda_{i\mu}^j(x,\vec{y})=h_\mu(x,\vec{y}){\epsilon_i}^j$.
The matrices $g_{\cD|_\tau i}^j(\vec{y})$ are
$\mbox{SO}(2)\cong \mbox{U}(1)$-matrices; they simply rotate
the fields $q_1$, $q_2$ into each other.
Finally we consider the particularly simple case of quantum mechanics
on a closed timelike curve. We set $d=1$, thus reducing the space
manifold $S$ to a single point. A closed curve $\alpha$ starting and ending
in $T_0\in S^1$ can, up to homotopy transformations, be classified
by its winding number $n\in \sZ$.
Let
\beq
g=\left(
  \begin{array}{cc}
  \cos\theta & \sin\theta \\
  -\sin\theta & \cos\theta
  \end{array}
  \right)
\eeq
be the $2\times 2$-matrix
corresponding to a parallel transport of $q_1$ and $q_2$ around a closed curve
with winding number $1$. The corresponding transformation for a curve with
winding number $n$ is $g^n$.
If $\theta=2\pi m$, $m\in\sZ$, then $g=\opone$.
If $\theta=2\pi q$, with $q$ being a rational number, but not an integer,
then the set $\{g^n\}$ is isomorphic to one of
the groups $\sZ_l$, $l\in \sN$, $l>1$.
If $\theta=2\pi r$, r being an arbitrary irrational number,
then the set $\{g^n\}$ is dense in $\mbox{U}(1)$.

\dgsection{Summary and Conclusions}
\noindent
In this paper we have developed a geometric formulation of quantum theory
on a space-time manifold. The evolution of the state vector is formulated
as a parallel transport along the observer's
worldline in a vector bundle over the space-time manifold.
A non-trivial holonomy group of the corresponding connection can
be viewed as a local symmetry group since the parallel transport of
state vectors (or observables) along closed loops should leave
expectation values invariant.

The evolution of the state vector is
given by a Schr\"{o}dinger equation
and therefore always unitary.
This may offer the possibility to formulate a consistent unitary
quantum field theory in the presence of closed timelike curves,
where the standard formulation fails \cite{4,5,6}.

Holonomies arise in the case of quantum systems in a different situation as
well. A quantum system can acquire a non-trivial Berry phase
\cite{7,8}
(or one of its non-Abelian generalizations, see
\cite{9,10}) if, in the case of adiabatic time evolution,
its parameters are changed from the outside.
In this case time evolution is described by a Hamiltonian
$H(\vec{\lambda}(t))$, where $\vec{\lambda}(t)$ denotes the set of
external time-dependent parameters.
Under certain circumstances, a Berry phase is an observable quantity.
In the theory developed in this paper the situation is different.
The objects describing the time evolution are the generators of the parallel
transport operators themselves. The non-trivial phase is acquired
if the point of description of the system is parallel-transported
along a closed curve in space-time, not in a parameter space. And, finally,
the phase is not observable. Precisely because of the latter property
the non-trivial holonomy gives rise to a symmetry.

Open questions are related to the following issues.
\begin{itemize}
\item[\itemsym]
We assumed that the generators $H_\mu$ of the parallel transport
operators and the representation $\varphi$
of the set of observables were given {\em a priori}. In order to study
realistic models, the $H_\mu$ should be derived from some basic principle.
It would certainly be desirable to make contact
with Lagrangian field theories\footnote{
It is even conceivable to formulate a `quantum gauge theory'
of the `gauge field' $H_\mu(x)$ and the
`matter fields' $F_i(x)$ by means of a `quantum action'
\beq
S=\int_M \di x \sqrt{g(x)}
\left[
-\frac{1}{4}\mbox{tr}\left(F_{\mu\nu}F^{\mu\nu}\right)
+ \mbox{tr}\left(D_\mu F_i(x) D^\mu F_i(x)\right)+\ldots
\right],
\eeq
where `$\mbox{tr}$' is the trace in the corresponding fibre of the
vector bundle $\pi_G$, $F_{\mu\nu}$ is the field strength
of $H_\mu$ and $D_\mu=\partial_\mu-H_\mu$
the covariant derivative.
$H_\mu$ and $F_i$ would then be the
solutions of the classical equations of motion derived from
the variational principle $\delta S=0$.
}.
\item[\itemsym]
The theory was formulated as a way of describing
a quantum field, say, in a background space-time.
A similar description, however,
is applicable for a quantum system located in a small region of space
being moved around in space-time.
The corresponding parallel transport operators are then active
transformations.
\item[\itemsym]
The measurement process
must also be formulated consistently. The possibility of a non-trivial
`twisted' bundle
$\pi_G:G\rightarrow M$
may be important here.
\item[\itemsym]
So far only scalar observables have been considered. The theory can
easily be extended to vector observables by considering bundles
$\pi_{G^\prime}:G^\prime\rightarrow TM$ over the tangent bundle of $M$.
Similarly tensor and spinor objects can be treated.
In the cases just mentioned a new structure has to be constructed:
that of state transformations at a point, connecting the state vectors
of observers whose coordinate systems are related by
general coordinate transformations.
\item[\itemsym]
A further complication arises if the Unruh effect is taken into account:
accelerated observers will register events even if the system is in the
vacuum state
\cite{11}. The formalism would have to be extended to mixed states, and the
set of states would be a bundle over the jet bundle of $M$, thus allowing
us to take into account higher derivatives of the
observer's worldline (i.e. the observer's acceleration).
\item[\itemsym]
We have interpreted the holonomy group as a group of local
symmetry transformations. A very interesting question is whether
this local symmetry can in general be related
to a gauge symmetry acting on internal degrees of freedom.
\item[\itemsym]
A final remark concerns the use of a classical background space-time.
A similar formulation of quantum theory should be possible
as soon as the `background structure' admits the
geometric objects used in this paper, i.e. bundles and parallel transport
operators.
This remark may apply to non-commutative structures
\cite{12} and to quantum spaces \cite{13},
but also to quantum sets
\cite{14} and to algebraic
and geometric lattices.
\end{itemize}

\vspace{0.5cm}
\noindent
{\Large\bf Acknowledgements}

\noindent
I wish to thank L.~Alvarez-Gaum\'{e}, R.~Haag and
E.~Verlinde for conversations
and for constructive
criticism. Moreover I am grateful to the organizers of the Ringberg workshop
in 1994 on `Space, Time and Quantum Theory' for creating a pleasant
and fruitful atmosphere for discussions and for the exchange of ideas
beyond standard concepts in fundamental physics.


\newpage
\newcommand{\scs}{\rm}
\newcommand{\bibitema}[1]{\bibitem[#1]{#1}}
\newcommand{\bibbeginlong}{\begin{thebibliography}{XXX\,1988\,\,}}
\newcommand{\bibbeginshort}{\begin{thebibliography}{XX}}

\bibbeginshort 

\bibitema{1}
   {\scs N.D.~Birrell, P.C.W.~Davies,}
   {\it Quantum Field Theory in Curved Space},
   Cambridge University Press, Cambridge (1982)

\bibitema{2}
   {\scs S.~Mandelstam,} Phys. Rev. \underline{D19} (1979) 2391

\bibitema{3}
   {\scs S.~Kobayashi, K.~Nomizu,}
   {\em Foundations of Differential Geometry},
   John Wiley \& Sons, New York (1963)

\bibitema{4}
   {\scs J.L.~Friedman, N.J.~Papastamatiou, J.J.~Simon,}
   Phys. Rev. \underline{D46} (1992) 4456

\bibitema{5}
   {\scs D.~Boulware,} Phys. Rev. \underline{D46} (1992) 4421

\bibitema{6}
   {\scs H.D.~Politzer,} Phys. Rev. \underline{D46} (1992) 4470

\bibitema{7}
   {\scs M.V.~Berry,} Proc. R. Soc. Lond. \underline{A392} (1984) 45

\bibitema{8}
   {\scs B.~Simon,} Phys. Rev. Lett. \underline{51} (1983) 2167

\bibitema{9}
   {\scs F.~Wilczek, A.~Zee,} Phys. Rev. Lett. \underline{52} (1984) 2111

\bibitema{10}
   {\scs G.~Papadopoulos,} J. Phys. \underline{A25} (1992) 2071

\bibitema{11}
   {\scs W.G.~Unruh,} Phys. Rev. \underline{D14} (1976) 870

\bibitema{12}
   {\scs R.~Coquereaux,}
   {\it Noncommutative Geometry and Theoretical Physics},
   CPT-88/P-2147 (1988)

\bibitema{13}
   {\scs J.~Wess, B.~Zumino,}
   Nucl. Phys. Proc. Suppl. \underline{18B} (1991) 302

\bibitema{14}
   {\scs D.~Finkelstein,}
   {\it Quantum Set Theory and Geometry},
   in {\it Quantum Theory and the Structures of Time and Space, Vol. 4},
   eds. L.~Castell, M.~Drieschner, C.F.~v.~Weizs\"{a}cker,
   Carl Hanser Verlag, M\"{u}nchen, Wien (1981)

\end{thebibliography}

\end{document}